\documentstyle[preprint, cite, aps]{revtex}

\begin{document}
\draft
\title{Relativistic Scalar Aharonov-Bohm Scattering}
\author{M. Gomes$^{\,a}$, J. M. C. Malbouisson$^{\,b}$, A. G. Rodrigues$^{a}$
and A. J. da Silva$^{\,a}$}
\address{$^{a\,}$Instituto de F\'{\i}sica, Universidade de S\~{a}o Paulo,
Caixa Postal 66318,\\
05315--970, S\~{a}o Paulo, SP, Brazil.}
\address{$^{b\,}$Instituto de F\'{\i}sica, Universidade Federal da Bahia,
Campus de Ondina,\\
40210--340, Salvador, BA, Brazil.}
\maketitle

\begin{abstract}
We discuss the scattering of relativistic spin zero particles by an
infinitely long and arbitrarily thin solenoid. The exact solution of the
first-quantized problem can be obtained as a mimic of the nonrelativistic
case, either in the original Aharonov-Bohm way or by using the Berry's
magnetization scheme. The perturbative treatment is developed in the
Feshbach-Villars two-component formalism for the Klein-Gordon equation and
it is shown that it also requires renormalization as in the Schr\"{o}dinger
counterpart. The results are compared with those of the field theoretical
approach which corresponds to the two-body sector of the scalar Chern-Simons
theory.
\end{abstract}

\section{INTRODUCTION}

The Aharonov-Bohm (AB) effect \cite{ab}, the scattering of a charged
particle by an infinitely long and arbitrarily thin solenoid, presents a
very peculiar situation of nonrelativistic (NR) quantum dynamics, with
charged particles feeling the vector potential in regions where the
electromagnetic field is null. It is an exactly solvable quantum mechanical
problem \cite{ab,berry} which, due to the singular nature of the potential,
requires the use of renormalization procedures to make its perturbative
treatment meaningful. In fact, as noticed by Corinaldesi and Rafeli \cite{cr}
, the bare perturbation theory leads to an incomplete result in the Born
approximation and to a divergent one in second order. For spinless particles
in quantum mechanics, the necessary renormalization is accomplished by
adding a delta function potential \cite{qm}, while in the second quantized
version it is implemented by introducing a $\phi ^4$ self-interaction, with
an appropriated strength, in a scalar Chern-Simons Lagrangian \cite
{hag2,jac,bl}.

In this paper, we discuss the relativistic scalar AB scattering, that is the
scattering of a relativistic charged spin zero particle by a thin fixed
solenoid in the viewpoint of the first quantization comparing with the
analysis in the framework of the field theory\cite{Boz,gms2,gms3}. In
section II, it is shown that the problem can be solved exactly through a
mimic of either the original AB solution or the Berry's magnetization
scheme. We then develop (section III) a perturbative analysis of the
Klein-Gordon equation in the presence of the solenoid using the
Feshbach-Villars two-component formalism and show that the bare perturbation
treatment presents similar problems as those of the nonrelativistic case and
possesses an analogous renormalization. In section IV, the results of the
field theoretical perturbative approach, corresponding to the two-body
sector of the Chern-Simons theory, are presented and compared with the
scattering amplitude obtained in the relativistic quantum mechanics.
Finally, some conclusions are outlined.

\section{EXACT SOLUTION IN THE FRAMEWORK OF THE FIRST QUANTIZATION}

The Klein-Gordon equation in the presence of an external electromagnetic
field, fixing the Coulomb gauge (${\bf \nabla \cdot A} = 0$), can be written
(in natural units, $\hbar =c=1$) as

\begin{equation}
\left( {\partial}^{2}_{t} - {\bf \nabla}^{2}+m^2+ {\cal U} \right) \phi = 0
\label{KG}
\end{equation}
where

\begin{equation}
{\cal U}=eA^{0}-i2e{\bf A\cdot \nabla }+e^{2}{\bf A}^{2}\;.
\label{potencial}
\end{equation}
For an ideal AB solenoid (a line carrying magnetic flux $\Phi $) at origin,
the magnetic field 
\begin{equation}
{\bf B}=\Phi \delta ({\bf r}){\hat{{\bf z}}\;,}  \label{B}
\end{equation}
with ${\bf r}=(x^{1},x^{2},0)$, and one may choose

\begin{equation}
A^{0}=A^{3}=0\;\;,\qquad A^{i}=-\frac{\Phi }{2\pi }\frac{\epsilon
^{\,i\,j}\,x^{\,j}}{{\bf r}^{2}}\;,\;i=1,2\;,
\end{equation}
where $\epsilon ^{\,i\,j}$ is the anti-symmetric symbol ($\epsilon
^{\,1\,2}=1$). The potential (\ref{potencial}) then becomes

\begin{equation}
{\cal U}_{Sol}=-i \left( \frac {e\Phi}{\pi} \right) \frac {{\bf r\times
\nabla}}{r^2}+ \left( \frac {e\Phi}{\pi} \right)^2 \frac 1{r^2}
\label{potsol}
\end{equation}
showing that the problem, owing to the symmetry, is actually a
two-dimensional one.

Considering a particle with (positive) energy $w_{{\bf p}}$ (${\bf p}$ in
the $x^1$ negative direction), the wave function can be separated as $\phi ( 
{\bf r},t)=\exp (-iw_{{\bf p}}t)\phi ({\bf r})$ and (\ref{KG}) reduces (in
cylindrical coordinates) to

\begin{equation}
\left[ \frac 1r\frac \partial {\partial r}\left( r\frac{\partial \phantom a}{
\partial r}\right) +\frac 1{r^2}\left( \frac \partial {\partial \theta }
+i\alpha \right) ^2+p^2\right] \phi (r,\theta )=0  \label{partwave}
\end{equation}
where $\alpha =e\Phi /2\pi $ is the magnetic flux parameter and $p^2=w_{{\bf 
p}}^2-m^2$. This equation coincides with the one solved by Aharonov and Bohm
for the nonrelativistic particle if one replaces the dispersion relation by $
p^2=2mE$, $E$ being the nonrelativistic energy. Thus, the exact solution of
( \ref{partwave}), vanishing when $r\rightarrow 0$ as required by the
``impenetrability'' condition, is given by \cite{ab}

\begin{equation}
\phi (r,\theta )=\sum_{l=-\infty }^{+\infty }(-i)^{|l-\alpha |}J_{|l-\alpha
|}(pr)\exp (il\theta )  \label{gensolu}
\end{equation}
where $J_{|l-\alpha |}$ denotes a Bessel function of first kind.

This exact solution of the Klein-Gordon equation in the presence of the
solenoid can also be obtained, in a rather distinct way, by applying the
Berry's magnetization scheme \cite{berry} to its solution in the absence of
the flux line, corresponding to the incident plane wave. Using the Fourier
expansion of a plane wave, 
\begin{equation}
\phi _0(r,\theta )=\exp \left[ -ipr\cos (\theta )\right] =\sum_{l=-\infty
}^{+\infty }(-i)^{|l|}J_{|l|}(pr)\exp (il\theta )\;,  \label{explane}
\end{equation}
and the Poisson summation formula 
\begin{equation}
\sum_{l=-\infty }^{+\infty }f(l)=\sum_{m=-\infty }^{+\infty }\int_{-\infty
}^{+\infty }d\eta f(\eta )\exp (2\pi im\eta )\;,  \label{Poissonsum}
\end{equation}
one obtains the whirling-wave expansion of the free solution 
\begin{eqnarray}
\phi _0(r,\theta ) &=&\sum_{m=-\infty }^{+\infty }w_m(r,\theta )\;, \\
w_m(r,\theta ) &=&\int_{-\infty }^{+\infty }d\eta \exp (-\frac 12i\pi |\eta
|)J_{|\eta |}(pr)\exp \left[ i\eta (\theta +2\pi m)\right] \;;  \label{wwexp}
\end{eqnarray}
notice that $w_m$ is not single-valued but satisfies $w_m(r,\theta +2\pi
)=w_{m+1}(\theta )$, therefore guaranteeing that $\phi _0(r,\theta )$ has
unique value.

The ingenious idea of Berry \cite{berry} to rescue the Dirac's magnetization
prescription, by which incorporating the phase factor $\exp \left( iq\int_{
{\bf r}_{0}}^{{\bf r}}{\bf A}({\bf r}^{\prime })\cdot d{\bf r}^{\prime
}\right) $ to the free wave function would lead to the solution in the
presence of the magnetic field, and to produce a single-valued wave function
was to apply the Dirac's procedure to each whirling-wave separately and to
resum the ``magnetized'' expansion. Doing so, one finds 
\begin{eqnarray}
\phi _{0}^{D}(r,\theta ) &=&\sum_{m=-\infty }^{+\infty }w_{m}^{D}(r,\theta
)\;, \\
w_{m}^{D}(r,\theta ) &=&\int_{-\infty }^{+\infty }d\eta \exp (-\frac{1}{2}
i\pi |\eta |)J_{|\eta |}(pr)\exp \left[ i(\eta +\alpha )(\theta +2\pi m) 
\right] \;,  \label{wwexpmag}
\end{eqnarray}
and making the change of variables $\eta ^{\prime }=\eta +\alpha $, the
inverse Poisson transform leads to the exact solution (\ref{gensolu}).
Notice that the Berry's procedure relies on the fact that the free solution
is a plane wave an so it can be applied equally well to both relativistic
and nonrelativistic Aharonov-Bohm scattering. An extension of the whirling
wave formulation of Berry to the case of the Dirac equation has been
presented recently \cite{gr}.

The AB scattering amplitude can be obtained by analyzing the asymptotic
behavior of the exact solution (\ref{gensolu}). It can be shown that \cite
{ab,cr} 
\begin{equation}
\phi (r,\theta )\stackrel{r\gg 1}{\longrightarrow }{\rm e}^{-ipr\cos \theta
}+{\rm e}^{i\pi /4}{\cal A}_{{\rm AB}}(|{\bf p}|,\theta )\frac{{\rm e}^{ipr} 
}{\sqrt{r}}  \label{asympt}
\end{equation}
where the scattering amplitude is given by 
\begin{equation}
{\cal A}_{{\rm AB}}(|{\bf p}|,\theta )\,\,=\,\,-\frac{i}{\sqrt{2\pi p}}\sin
(\pi \alpha )\,\,\left[ \tan \left( \frac{\theta }{2}\right) -i\,{\rm sgn\,}
(\alpha )\right] \;,  \label{AB}
\end{equation}
with ${\rm sgn\,}(\alpha )=|\alpha |/\alpha \,$. It should be remarked that
the exact solution, and consequently the AB scattering amplitude, can be
also obtained within the two-component formalism, which is employed to
construct the perturbative expansion of the $S$ matrix in the next section.

\section{PERTURBATIVE ANALYSIS OF THE RELATIVISTIC AB SCATTERING}

In order to be able to use the standard perturbation theory one has to cast
the Klein-Gordon equation as a differential equation of first order in time,
that is in a form similar to the Schr\"odinger equation. This can be done
following the prescription below.\ 

\subsection{Two-component formalism of the Klein-Gordon equation}

In the Feshbach-Villars representation \cite{fv}, the Klein-Gordon wave
function in the presence of an external field $A^\mu $ is written in the form

\begin{equation}
\Psi =\left( 
\begin{array}{c}
\chi \\ 
\zeta
\end{array}
\right) \;\;\;\longleftrightarrow \;\;\;\left\{ 
\begin{array}{c}
\chi =\frac{1}{\sqrt{2m}}\left[ i\partial _{t}\phi -\left( eA^{0}-m\right)
\phi \right] \\ 
\;\,\,\zeta =\frac{1}{\sqrt{2m}}\left[ -i\partial _{t}\phi +\left(
eA^{0}+m\right) \phi \right]
\end{array}
\right. \;\;.
\end{equation}
This two-component wave function satisfies a Schr\"{o}dinger like equation,

\begin{equation}
\,i\partial _{t}\Psi =H\Psi \;,
\end{equation}

\noindent with the ``Hamiltonian'' given by

\begin{equation}
H=\left( m-\frac{({\bf \nabla }-ie{\bf A})^2}{2m}\right) \tau _3-\frac{({\bf 
\nabla }-ie{\bf A})^2}{2m}\,i\tau _2+e\,A^0
\end{equation}
where $\tau _i$ are the Pauli matrices. Although $H\,$ is not Hermitian, $
\tau _3H^{\dag }\tau _3=H$ and the norm $\int d{\bf x}\Psi ^{\dagger }\,\tau
_3\,\Psi \,$ is conserved, i. e., $\tau _3$ plays the role of a metric
tensor.

This formalism allows the use of perturbation theory in the standard way and
we make the partition $\,H=H_0+H_{{\rm int}}=H_0+V+H_1\,\left[ \tau _3+i\tau
_2\right] \,$ where $V=e\,A^0\,$ and

\begin{eqnarray}
H_{0} &=&\left( m-\frac{{\bf \nabla }^{2}}{2m}\right) \tau _{3}-\frac{{\bf 
\nabla }^{2}}{2m}\,i\tau _{2}\;, \\
H_{1} &=&\left( \frac{ie}{2m}\,\left[ {\bf \nabla \cdot A}+{\bf A\cdot
\nabla }\right] +\frac{e^{2}}{2m}\,{\bf A}^{2}\right) \;.
\end{eqnarray}
The free positive and negative energy solutions are given by 
\begin{equation}
\Psi _{{\bf p}}^{(+)}({\bf x},t)={\rm e}^{-iw_{{\bf p}}t}\,\Psi _{{\bf p}
}^{(+)}({\bf x})=\frac{1}{{2\pi }}\frac{w_{{\bf p}}+m}{\sqrt{4mw_{{\bf p}}}}
\left( 
\begin{array}{c}
1 \\ 
\frac{m-w_{{\bf p}}}{m+w_{{\bf p}}}
\end{array}
\right) {\rm e}^{\,i{\bf p}\cdot {\bf x}-iw_{{\bf p}}t}\;,  \label{possol}
\end{equation}
\begin{equation}
\Psi _{{\bf p}}^{(-)}({\bf x},t)={\rm e}^{+iw_{{\bf p}}t}\,\Psi _{{\bf p}
}^{(-)}({\bf x})=\frac{1}{{2\pi }}\frac{w_{{\bf p}}+m}{\sqrt{4mw_{{\bf p}}}}
\left( 
\begin{array}{c}
\frac{m-w_{{\bf p}}}{m+w_{{\bf p}}} \\ 
1
\end{array}
\right) {\rm e}^{\,-i{\bf p}\cdot {\bf x}+iw_{{\bf p}}t}\;,
\end{equation}
which satisfy the normalization conditions 
\begin{equation}
\int d{\bf x}\Psi _{{\bf p^{\prime }}}^{(\pm )}\tau _{3}\Psi _{{\bf p}
}^{(\pm )}=\pm \delta ({\bf p^{\prime }}-{\bf p})\;.
\end{equation}
Notice that the Berry's magnetization procedure can be applied to the
positive energy solution (\ref{possol}) leading to the exact solution
described in the last section and thus to the AB scattering amplitude (\ref
{AB}).

In fact, the two-component formalism is the appropriate scenario to
implement the perturbative treatment of the scalar AB scattering in the
framework of the relativistic quantum mechanics. The $S$ matrix elements can
be calculated through the formula 
\begin{equation}
S_{fi}=<f|{\rm T}\exp \left[ -i\int dtH_{I}(t)\right] |i>\;,  \label{Smatrix}
\end{equation}
where {\rm T} stands for the time ordering operator and $H_{I}(t)$ is the
interaction picture of the interaction Hamiltonian $H_{{\rm int}}$, from
which we can define the scattering amplitude ${\cal A}_{fi}$ as 
\begin{equation}
S_{fi}=-2\pi i\delta (w_{f}-w_{i}){\cal A}_{fi}\;.
\end{equation}

\subsection{The bare scattering amplitude}

The S matrix, expressing the scattering of a positive energy particle, in
the first Born approximation, becomes

\begin{equation}
S_{fi}^{(1)}=-i\int dt\,d{\bf x}\,\Psi _{{\bf p}^{^{\prime }}}^{\dagger }( 
{\bf x})\,\tau _3\,H_{{\rm int}}({\bf x,t)\,}\Psi _{{\bf p}}({\bf x}
)=-i\,2\pi \,\delta (w_{{\bf p}^{^{\prime }}}-w_{{\bf p}})\, {\cal A}
_{fi}^{(1)}\;\;
\end{equation}
with the reduced amplitude given by

\begin{equation}
{\cal A}_{fi}^{(1)}=\int d{\bf x\,}{\rm e}^{\,-i\,{\bf p}^{^{\prime }}{\bf 
\cdot \,x}}\,\left[ \frac{m}{w_{{\bf p}}}\,H_{1}({\bf x},{\bf \nabla })+V( 
{\bf x)}\right] \,{\rm e}^{\,i\,{\bf p\,\cdot \,x}}\;.
\end{equation}

\noindent Considering the Coulomb gauge, ${\bf \nabla \cdot A}=0\,$, and for
the moment neglecting the higher order term $\,e^2{\bf A}^2$, one obtains

\begin{equation}
{\cal A}_{fi}=\int d{\bf x\,}\left[ -\frac e{w_{{\bf p}}}{\bf A}({\bf x})\, 
{\bf \cdot \,p}+V({\bf x})\right] \,{\rm e}^{-\,i\,({\bf p}^{\prime }-{\bf p}
)\,{\bf \cdot \,x}}=-\frac e{w_{{\bf p}}}\,{\bf \tilde A(q})\,{\bf \cdot \,p}
+\tilde V({\bf q})\;\;\;
\end{equation}
where ${\bf q=p}^{\prime }-{\bf p}\,$.

For the AB potential, $V=0\, $ and

\begin{equation}
\tilde{A}^{\,i}({\bf q})=\frac{i\,\Phi }{2\pi }\,\epsilon
^{\,i\,j}\lim_{\lambda \rightarrow 0}\partial _{q^{\,j}}\int d{\bf r\,}\frac{
\,{\rm e}^{\,i\,{\bf q}\,{\bf \cdot \,r}}}{{\bf r}^{2}+\lambda ^{2}}\,=i\Phi
\,\epsilon ^{\,i\,j}\lim_{\lambda \rightarrow 0}\,\partial
_{q^{\,j}}\,K_{0}(\,\lambda \,|{\bf q}|\,)=-\,i\,\Phi \,\frac{\epsilon
^{\,i\,j}\,q^{\,j}}{{\bf q}^{2}}\;,  \label{FourierA}
\end{equation}
$K_{0}$ denoting the modified Bessel function, and therefore the reduced
amplitude is given by

\begin{equation}
{\cal A}_{fi}^{(1)}=-\frac{i\,e\,\Phi }{w_{{\bf p}}}\,\frac{{\bf p\times q}}{
{\bf q}^2}=-\frac{i\,e\,\Phi }{2\,w_{{\bf p}}}\,\cot \left( \frac{\theta _S}2
\right)
\end{equation}
where $\theta _S$ is the scattering angle. Notice, since $\theta _S$ is the
angle between the incoming (${\bf p}$) and the outgoing (${\bf p}^{\prime }$
) momenta, it is related to the AB polar angle by $\theta _S=\pi -\theta $.
One immediately notices the absence of the nonanalytic term occurring in the
first order of the expansion of the exact result (\ref{AB}).

To calculate higher-order contributions one should be aware of the fact
that, since the interaction with the external field contains terms of order $
e$ and $e^2$, the perturbative expansion in the coupling constant does not
correspond ``order by order'' to the Born expansion. Thus, to find the
second order contribution to the amplitude, besides the contribution coming
from the second term of the expansion (\ref{Smatrix}), one has to account
for the ${\bf A}^2$ term present in the first Born approximation; one obtains

\begin{equation}
{\cal A}_{fi({\bf A}^{2})}^{(2)}=\int d{\bf r\,}{\rm e}^{-i{\bf p}^{\prime
}\cdot {\bf r}}\left[ \frac{m}{w_{{\bf p}}}\frac{e^{2}}{2m}{\bf A}^{2}({\bf 
r })\right] {\rm e}^{i{\bf p}\cdot {\bf r}}=\frac{m}{w_{{\bf p}}}\frac{e^{2} 
}{ 2m}\,\tilde{{\cal C}}({\bf q})
\end{equation}

\noindent where, the Fourier transform of ${\bf A}^{2}$, 
\begin{equation}
\tilde{{\cal C}}({\bf q})=\left( \frac{\Phi }{2\pi }\right) ^{2}\int d{\bf r}
\frac{{\rm e}^{-i{\bf q\cdot r}}}{r^{2}}=\left( \frac{\Phi }{2\pi }\right)
^{2}\lim_{\lambda \rightarrow 0}\int d{\bf r}\frac{{\rm e}^{-i{\bf q\cdot r}
} }{r^{2}+\lambda ^{2}}=\left( \frac{\Phi }{2\pi }\right) ^{2}\lim_{\lambda
\rightarrow 0}K_{0}(\lambda |{\bf q}|)\;,
\end{equation}

\noindent leads to a divergent contribution. Notice that, it is precisely
due to the absence of the ${\bf A}^2$ term in the interaction of the spin
half particles with the field that no divergence occurs in that case;
furthermore the Pauli magnetic interaction makes naturally the first Born
approximation correct\cite{Vera}. Wishing to be able to suppress the
divergence of the scalar case, we keep the regularized contribution

\begin{equation}
{\cal A}_{fi\,\,({\bf A}^{2})}^{(2)\,\,Reg}=\frac{e^{2}\Phi ^{2}}{4(2\pi
)^{3}}\frac{1}{w_{p}}\left\{ -\ln (\lambda ^{2})-\ln (p^{2})-\ln [2(1-\cos
\theta _{S})]+2(\ln 2-\gamma )\right\} \;.
\end{equation}
where $\gamma $ is the Euler's constant.

The other second order contribution comes from the second Born
approximation, 
\begin{equation}
S_{fi}^{(2)}=-\int \int dt_{1}dt_{2}\,\theta (t_{2}-t_{1})\int d{\bf r_{2}}
\,\Psi _{{\bf p^{\prime }}}^{(+)\,\dagger }({\bf r_{2}})\,\tau
_{3}\,H_{int}( {\bf r_{2}},t_{2})\,H_{int}({\bf r_{2}},t_{1})\,\Psi _{{\bf p}
}^{(+)}({\bf r_{2}})\;,
\end{equation}
considering only the part of the interaction Hamiltonian which is first
order in $e$. Using the completeness relation 
\begin{equation}
\int d{\bf k}\left[ \,\Psi _{{\bf k}}^{(+)}({\bf r_{1}})\,\Psi _{{\bf k}
}^{(+)\,\dagger }({\bf r_{2}})\,\tau _{3}-\Psi _{{\bf k}}^{(-)}({\bf r_{1}}
)\,\Psi _{{\bf k}}^{(-)\,\dagger }({\bf r_{2}})\,\tau _{3}\,\right] ={\bf I}
\,\,\delta ({\bf r_{1}-r_{2}})
\end{equation}
and the identity 
\begin{equation}
\theta (t)E^{n}{\rm e}^{-iEt}=\frac{1}{2\pi i}\int d\omega \frac{\omega ^{n} 
{\rm e}^{-i\omega t}}{E-\omega -i\epsilon }\;,
\end{equation}
one gets 
\begin{equation}
S_{fi\,({\bf A\cdot p})}^{(2)}=-2\pi i\delta (w_{f}-w_{i}){\cal A}_{fi\,( 
{\bf A\cdot p})}^{(2)}
\end{equation}
with 
\begin{eqnarray}
{\cal A}_{fi\,({\bf A}\cdot {\bf p})}^{(2)} &=&-\frac{e^{2}}{(2\pi )^{4}} 
\frac{1}{w_{p}}\int \frac{d{\bf k}}{\omega _{k}}\left\{ \frac{[{\bf k}\cdot {
\ \tilde{{\bf A}}}({\bf p}^{\prime }-{\bf k})][{\bf p}\cdot {\tilde{{\bf A}}}
( {\bf k}-{\bf p})]}{\omega _{p}-\omega _{k}+i\epsilon }\right.  \nonumber \\
\phantom a &&\phantom{aaaaaaaaaaaaaaa}-\left. \frac{[{\bf k}\cdot \tilde{
{\bf A}}({\bf p}^{\prime }+{\bf k})][{\bf p}\cdot \tilde{{\bf A}}({\bf k}+ 
{\bf p})]}{\omega _{p}+\omega _{k}+i\epsilon }\right\} \;.  \label{A2AP}
\end{eqnarray}

\noindent where $w_{k}=\sqrt{{\bf k}^{2}+m^{2}}$. From (\ref{FourierA}), it
follows that ${\bf k}\cdot \tilde{{\bf A}}({\bf p})=i\Phi ({\bf k}\times 
{\bf p})/p^{2}$, so the contribution (\ref{A2AP}) to the second order
amplitude is given by 
\begin{eqnarray}
{\cal A}_{fi\,({\bf A}\cdot {\bf p})}^{(2)} &=&\frac{e^{2}\Phi ^{2}}{(2\pi
)^{4}}\frac{1}{w_{p}}\int d{\bf k}\frac{({\bf k}\times {\bf p}^{\prime })( 
{\bf k}\times {\bf p})}{({\bf k}-{\bf p}^{\prime })^{2}({\bf k}-{\bf p})^{2}}
\frac{2}{p^{2}-k^{2}+i\epsilon }  \nonumber \\
\ &=&\frac{e^{2}\Phi ^{2}}{4(2\pi )^{3}}\frac{1}{w_{p}}\left\{ \ln [2(1-\cos
\theta _{S})]+i\pi \right\} \;,
\end{eqnarray}
where the integral was performed as presented in \cite{gms3}. The $\theta
_{S}$ dependent part of this finite contribution is canceled out with one of
the terms of $\,{\cal A}_{fi({\bf A}^{2})}^{(2)}$. One thus obtains, adding
up both contributions, a divergent second order term for the bare scattering
amplitude. To recover a finite result, coinciding with the expansion of the
exact amplitude, one has to implement some renormalization procedure.

\subsection{Perturbative renormalization}

To regain the correct perturbative expansion we have to search for an
appropriate counterterm, an additional interaction, which should suppress
the divergence of the second order and contributes in the lowest order
recovering the exact result. At first sight, one can try to follow the case
of spin half particles and look for a magnetic type of interaction to do the
job. However, in the scalar case the magnetic interaction, given by $
H_{mag}=g\epsilon ^{\mu \nu \rho }F_{\nu \rho }\,j_{\mu }$, where $F_{\nu
\rho }$ is the field strength and $j_{\mu }$ the particle's current, can
only furnish the correct result in the leading nonrelativistic order\cite
{Zimerman}, and thus can not describe the full relativistic case. One of the
reasons for that is the fact that such interaction does not have the same
matrix structure in the two-component formalism as the ${\bf A}^{2}$ term,
which is responsible for the divergence one wishes to eliminate.

The simplest additional interaction having the same matrix structure of the
magnetic potential term one can consider, leading to the same logarithmic
divergence in second order, is a pure delta function external potential. In
fact, by adding to (\ref{potencial}) the term 
\begin{equation}
{\cal U}_{(delta)}=g\delta ({\bf r})
\end{equation}
one obtains, in first order, the contribution 
\begin{equation}
{\cal A}_{fi(delta)}^{(1)}=\frac{mg}2\frac 1{w_p}
\end{equation}
while the cutoff regularized ``delta-delta'' contribution to the second
order becomes 
\begin{eqnarray}
{\cal A}_{fi(delta)}^{(2)\,Reg} &=&\frac{m^2g^2}{(2\pi )^4}\frac 1{w_p}
\int^{\Lambda ^2}d{\bf k}\frac 1{w_p-w_k+i\epsilon }  \nonumber \\
\ &=&\frac{m^2g^2}{4(2\pi )^3}\frac 1{w_p}\left\{ -\ln (\Lambda ^2)+\ln
(p^2)-i\pi \right\}
\end{eqnarray}
The crossing terms involving the delta and ${\bf A}\cdot {\bf p}$
interactions need not to be considered; the sum of their contributions
vanishes, a fact which can be inferred from symmetry arguments.

Including these contributions from the delta potential, the first and the
regularized second order parts of the scattering amplitude become 
\begin{equation}
{\cal A}_{fi}^{(1)}=-i\frac{e\Phi }{2w_p}\cot \left( \frac{\theta _S}2
\right) +\frac{mg}{2w_p}
\end{equation}
\begin{eqnarray}
{\cal A}_{fi}^{(2)} &=&\frac 1{4(2\pi )^3}\frac 1w_p(m^2g^2-e^2\Phi ^2)[\ln
(p^2)-i\pi ]  \nonumber \\
\phantom a &&+\frac 1{4(2\pi )^3}\frac 1w_p\left[ -m^2g^2\ln (\Lambda
^2)-e^2\Phi ^2(\ln (\lambda ^2)-2\ln 2+2\gamma )\right]
\end{eqnarray}
One sees then that the agreement with the expansion of the exact result can
be reached if the strength of the delta interaction is fixed satisfying

\begin{equation}
m^2g^2= e^2\Phi^2= 4 \pi^2\alpha^2
\end{equation}

\noindent and the cutoffs are adjusted so that 
\begin{equation}
\Lambda \lambda =2\exp (-\gamma )\;;
\end{equation}
in doing so, the first order term, multiplied by the appropriate kinematical
factor, reproduces the correct result and the second order, proportional to $
\alpha ^{2}$, vanishes as it should.

\section{FIELD THEORETICAL APPROACH}

The scalar nonrelativistic AB scattering corresponds, in the field
theoretical approach, to the two-body sector of the theory of a Chern-Simons
field coupled with a self-interacting scalar field for which the Lagrangian
density is given by\cite{bl}

\begin{equation}
{\cal L}_{NR}=\psi ^{*}\left( iD_t+\frac{{\bf D}^2}{2m}\right) \psi -\frac{
v_0}4(\psi ^{*}\psi )^2+\frac \Theta 2\partial _t{\bf A}\times {\bf A}
-\Theta A_0{\bf \nabla }\times {\bf A}\;,  \label{lagranNR}
\end{equation}

\noindent where $D_{t}=\partial _{t}+ieA_{0}$ and ${\bf D}={\bf \nabla }-ie 
{\bf A}$ are covariant derivatives, $\Theta $ is the Chern-Simons parameter
and $v_{0}$ is the bare self-coupling. The renormalized nonrelativistic two
particle scattering amplitude, in the center of mass (CM) frame, is given,
up to one loop, by

\begin{equation}
{\cal A}_{NR}=-v-i\frac{2e^2}{m\Theta }\cot \theta +\frac m{8\pi }\left(
v^2- \frac{4e^4}{m^2\Theta ^2}\right) \left[ \ln \left( \frac{\mu ^2}{{\bf p}
^2} \right) +i\pi \right] \;,  \label{ANRren}
\end{equation}
where $\mu $ is an arbitrary mass scale that breaks the scale invariance of
the amplitude. By choosing the critical value $v_c^{+}=+2e^2/m|\Theta |\,$,
which corresponds to a repulsive quartic interaction, this amplitude reduces
to the first order ($e^2$) Aharonov-Bohm amplitude for identical particles
which is given by \cite{bl}

\begin{equation}
{\cal F}_{{\rm AB}}(|{\bf p}|,\theta )=-i\frac{4\pi \beta }m\,\,\left[ \cot
\theta _S-i\,{\rm sgn\,}(\beta )\right] \;,  \label{ABC}
\end{equation}
where $\beta =e^2/2\pi \Theta $ coincides with the flux parameter $\alpha $
if the identification $\Theta =e/\Phi $ is made; this symmetrized amplitude
is the same, adjusting the kinematical factor, as the one obtained from the
first order expansion of the exact NR result (\ref{AB}).

It has been shown that the relativistic Chern-Simons theory, 
\begin{equation}
{\cal L}=(D_{\mu }\phi )^{\ast }(D^{\mu }\phi )-m^{2}\phi ^{\ast }\phi - 
\frac{\lambda }{4}(\phi ^{\ast }\phi )^{2}+\frac{\Theta }{2}\epsilon
_{\sigma \mu \nu }A^{\sigma }\partial ^{\mu }A^{\nu }\;,  \label{lagran}
\end{equation}

\noindent reduces to the nonrelativistic case in the leading approximation 
\cite{Boz}. We have calculated in this theory the $|{\bf p}|/m$ expansion of
the two particle amplitude, up to one loop order, using an intermediate
cutoff procedure introduced in ref. \cite{gms1}. The renormalized CM
amplitude, including the factor $1/4w_{{\bf p}}^2\,$ (where $w_{{\bf p}}= 
\sqrt{m^2+{\bf p}^2}\,$) which makes the states to have the same
normalization as in the NR case, can be written as ${\cal A} ={\cal A}
^{(0)}+ {\cal A}^{(1)}$ \cite{gms2}, where

\begin{equation}
{\cal A}^{(0)}=-\frac{\lambda }{4w_{p}^{2}}-\frac{ie^{2}}{\Theta w_{p}}\cot
\left( \frac{\theta _{S}}{2}\right) \,+\,\left[ \theta _{S}\rightarrow
\theta _{S}-\pi \right]  \label{AtreeNR}
\end{equation}
is the exact tree level contribution and the one loop term, up to order $
{\bf p}^{2}/m^{2}\,$, is given by

\begin{eqnarray}
{\cal A}^{(1)} &\simeq &\frac m{8\pi }\left( \frac{\lambda ^2}{16m^4}-\frac{
4e^4}{m^2\Theta ^2}\right) \left[ \ln \left( \frac{4m^2}{{\bf p}^2}\right)
+i\pi \right]  \nonumber \\
&&\ \ \ \ -\frac m{8\pi }\left( \frac{3\lambda ^2}{32m^4}-\frac{2e^4}{
m^2\Theta ^2}\right) \frac{{\bf p}^2}{m^2}\left[ \ln \left( \frac{4m^2}{{\bf 
p}^2}\right) +i\pi \right]  \nonumber \\
&&\ \ \ \ +\frac m{8\pi }\left( \frac{\lambda ^2}{4m^4}-\frac{14e^4}{
3m^2\Theta ^2}\right) -\frac m{8\pi }\left( \frac{25\lambda ^2}{96m^4}+\frac{
74e^4}{15m^2\Theta ^2}\right) \frac{{\bf p}^2}{m^2}\;.  \label{AtotalNR}
\end{eqnarray}
One sees that the leading term of the $|{\bf p}|/m$ expansion of ${\cal A}$
coincides with ${\cal A}_{NR}$ if one identifies $v=\lambda /4m^2$ and
choose $\mu ^2=4m^2\,$. Independently of the fixing of $\mu \,$, by taking $
\lambda _c^{+}=4m^2v_c^{+}\,$ the one loop contribution for the leading
order (in $|{\bf p}|/m$) vanishes and the tree level one reproduces the AB
scattering. However, the subdominant terms do not vanish for $\lambda
=\lambda _c^{+}$ and constitute additional relativistic corrections to the
AB effect \cite{gms2} which are originated from field theoretical effects as
vacuum polarization and vertex radiative corrections. Notice, in this
respect, that the vectorial interaction vertex expressing the bare coupling
between the matter and the Chern-Simon fields possesses, in the relativistic
case, an energy factor in the zeroth component which is not present in the
NR Lagrangian.

\section{CONCLUSIONS}

Using a two-component formalism, in this work we have studied perturbatively
the first quantized AB scattering of relativistic scalar particles. We
proved that, to eliminate divergences due to the $A^{2}$ coupling, it is
necessary to add a contact delta interaction which, in a field theoretical
language, corresponds to a $(\phi ^{\dagger }\phi )^{2}$ self-interaction.
Now, it is known that in the fermionic case Pauli's magnetic interaction $
\sim {\bf B}\cdot {\bf \,}\psi ^{\dagger }{\bf s}\psi $, with ${\bf s}$
standing for the spin operator, provides the necessary ingredient to make
the final result well defined. This immediately suggest that in the scalar
case the linearized form $\sigma (\phi ^{\dag }\phi )$, where in lowest
order the external field $\sigma =g\delta (x)$, should be added to the
original Lagrangian, as we did. In the nonrelativistic case the purely
magnetic coupling $g\epsilon ^{\mu \nu \rho }F_{\nu \rho }\,j_{\mu }$, where 
$F_{\nu \rho }$ is the field strength and $j_{\mu }$ the particle's current,
equally provides the cancellation of the divergence, but in the relativistic
domain it has to be disregarded since it does not have the appropriated
momentum dependence and, in the field theory context, it is
nonrenormalizable. Comparison between the first-quantized and the field
theoretical perturbative expansions shows that the latter has additional
contributions coming from vacuum polarization and vertex radiative
corrections. These terms, absent in a direct nonrelativistic approach, show
that the original Aharonov-Bohm problem is an idealized situation since
vacuum polarization makes the magnetic field necessarily nonvanishing
outside the solenoid.

\ 

{\bf Acknowledgments}

\thinspace

This work was partially supported by CAPES, CNPq and FAPESP, Brazilian
agencies.

\

\end{document}